# Quantum memory effects on the dynamics of electrons in small gold clusters

Yair Kurzweil[1] and Roi Baer[1,2]♦

[1]*Department of Physical Chemistry and the Lise Meitner Center for Quantum Chemistry, the Hebrew University of Jerusalem, Jerusalem 91904 Israel.*

[2]*Dept. of Chemistry and Biochemistry, University of California at Los Angeles CA 90095*

Electron dynamics in metallic clusters are examined using a time-dependent density functional theory that includes a "memory term", i.e. attempts to describe temporal non-local correlations. Using the Iwamoto, Gross and Kohn exchange-correlation (XC) kernel we construct a translationally invariant memory action from which an XC potential is derived that is translationally covariant and exerts zero net force on the electrons. An efficient and stable numerical method to solve the resulting Kohn-Sham equations is presented. Using this framework, we study memory effects on electron dynamics in spherical Jellium "gold clusters". We find memory significantly broadens the surface plasmon absorption line, yet considerably less than measured in real gold clusters, attributed to the inadequacy of the Jellium model. Two-dimensional pump-probe spectroscopy is used to study the temporal decay profile of the plasmon, finding a fast decay followed by slower tail. Finally, we examine memory effects on high harmonic generation, finding memory narrows emission lines.

## I. INTRODUCTION

Time dependent density functional theory[1] (TDDFT) is an in-principle exact theory of quantum many-body dynamics which does not require use of many-body wave functions. It thus forms a basis for an approximate, computationally tractable method for describing the dynamics of electrons in large molecules and nanostructures under the influence of electromagnetic fields. Any practical implementation of TDDFT must use severe approximations. As it turns out though, even simple approximations often yield reasonably accurate results. While these are not of spectroscopic accuracy, they are often useful for many purposes, and are comparable to the quality of computations which use many-body wave functions.

We will consider a system of $N_e$ electrons in their ground-state, which at time $t > 0$ are subject to a time dependent perturbation. TDDFT maps such a system of interacting electrons subject to an external potential $v_{ext}(\mathbf{R},t)$ onto a system of *non-interacting* identical fermions of the same mass which starts from its ground-state as well. The mapping is uniquely characterized by the fact that both systems have the same 1-particle density $n(\mathbf{R},t)$. The dynamics of the non-interacting fermions is determined by an external potential called the Kohn-Sham potential $v_{KS}(\mathbf{R},t)$, a complicated unknown functional of the density $n(\mathbf{R}',t')$. Since $v_{KS}(\mathbf{R},t)$ is an unknown functional, one builds into as much general physics as possible, isolating the unknown part of $v_{KS}$ into a hopefully small potential, which can be reasonably approximated. This latter potential is the exchange-correlation (XC) potential, defined by

$$v_{XC}(\mathbf{R},t) = v_{KS}(\mathbf{R},t) - v_{ext}(\mathbf{R},t) - v_H(\mathbf{R},t), \quad (1.1)$$

where $v_H(\mathbf{R},t) = \int n(\mathbf{R}',t)/|\mathbf{R}-\mathbf{R}'| d^3R'$ is the Hartree potential, describing the instantaneous classical Coulomb interaction. It is important to observe that the XC potential is a *universal* functional of the density, completely independent of the external potential. Thus, in actual applications, it is approximated using crude, simplified, but still universal functionals.

The simplest universal XC potential, called the adiabatic local density approximation (ALDA)[2] is based on the static properties of the homogeneous electron gas. ALDA is exact in the limit of low frequency/long wavelength variations in the density. The ALDA and other adiabatic functionals do not allow for explicit non-temporally local effects of the density on the XC potential. In principle, the XC potential at time $t$ should depend to some extent on the past history, i.e. on $n(\mathbf{r}',t')$ with $t' < t$. Non-adiabatic functionals are thus also called "memory" functionals. To date, most applications of time-dependent density functional theory are made using adiabatic, memoryless, functionals[3-17]. While some of the results are surprisingly good in view of the crude assumptions, there are known problems which prevent TDDFT from achieving high reliability. Some shortcomings are apparent even in the linear response regime. These include the inability to correctly describe excitations with multiple electron character[18] and (the related) grossly spurious prediction of plasmon decay in metal clusters[19]. It is reasonable to expect that in many strong-field situations, the problems associated with the adiabatic functionals will only get worse. Indeed electron ionization rates from a model of the He atom are not described well by ALDA[20].

Going beyond the adiabatic approximation and including memory effects is an important goal discussed extensively for some time[21-26]. However, there has been very slow progress in developing new memory functionals that can actually be applied to real systems. The problem hindering the development of memory functionals in TDDFT has to do with the imposition of some exact conditions compelling the behavior of universal XC potentials. One such condition is the translational covariance: if we translate the density by some time-dependent bector $\mathbf{x}(t)$, i.e.:

$$n'(\mathbf{R}',t) = n(\mathbf{R}' + \mathbf{x}(t), t). \quad (1.2)$$

The XC potential must rigidly follow:

---

♦ Corresponding author: FAX: +972-2-6513742, roi.baer@ huji.ac.il

$$v_{XC}[n'](\mathbf{R}',t) = v_{XC}[n](\mathbf{R}' + \mathbf{x}(t), t), \qquad (1.3)$$

This condition is natural and it was proved by Vignale[27], along with another physically obvious constraint, we call the zero force condition. Here one considers the electronic center of mass

$$\mathbf{R}_{CM}(t) = \frac{1}{N_e} \int n(\mathbf{R},t) \mathbf{R} d^3 R \qquad (1.4)$$

which must be identical for both interacting and non interacting systems. By Ehrenfest's law, $m_e \ddot{\mathbf{R}}_{CM} = \mathbf{F}$, where $\mathbf{F} = -\int \nabla v_{ext}(\mathbf{R}) n(\mathbf{R},t) d^3 R$ for the interacting system and $\mathbf{F} = -\int \nabla v_{KS}(\mathbf{R}) n(\mathbf{R},t) d^3 R$ for the non-interacting system. Thus both forces are equal to each other. Since the total Hartree force is always zero, it is concluded that the XC force vanishes:

$$\int [-\nabla v_{XC}(\mathbf{R},t)] n(\mathbf{R},t) d^3 R = 0. \qquad (1.5)$$

An analogous rule, concerning the total XC torque, can be proved only if the current density in the interacting and non-interacting systems is the same. However, this is not assured in TDDFT. In fact imposing such identity will carry us into the realm of time-dependent *current density functional theory*[28] (TDCDFT), which we avoid in this present work, for reasons we discuss now. While TDCDFT functionals have been developed and successfully applied[24, 25, 29-33], they are presently limited to linear response and slow density variations. Extending this approach to real-time and beyond linear response applications is complicated and has been done only in 1-dimensional cases[26, 33]. The basic problem is that imposition of TC usually warrants a Lagrangian system of coordinates[23, 26, 28, 34], as opposed to the normally used Eularian (fixed) coordinate system. In three dimensions, this is a great impediment since the numerical methods for solving the Schrödinger equation in a Lagrangian frame are still not developed and robust enough to serve a basis for a general TDDFT program. The problem is that the Lagrangian coordinate system relies on the electron velocity field $\mathbf{u}(\mathbf{r},t) = \mathbf{j}(\mathbf{r},t)/n(\mathbf{r},t)$ [where $\mathbf{j}(\mathbf{r},t)$ is the current density] which is numerically ill-defined whenever the density tends to zero.

In this paper we develop a practical method for studying numerically the memory effects in electron dynamics. We avoid Lagrangian frames and make use of a different, much simplified approach to memory functionals, following the suggestions of Vignale[27]. We define a family of translationally invariant (TI) "actions" on the Keldysh contour, which are 1) based on the simple linear response kernel of the homogeneous electron gas 2) are translationally invariant (a notion we describe in the next section). From each of these actions an XC potential $v_{XC}(\mathbf{r},t)$ can be derived which is 1) TC [Eq. (1.3)], and 2) obeys the zero force condition [Eq. (1.5)] and 3) is *causal* [ $v_{XC}(\mathbf{r},t)$ depends on $n(\mathbf{r}',t')$ where $t' < t$ ]. This approach we take can be considered a simple approximation to the full but difficult Lagrangian description. As such it is numerically doable, and is stable, as we find in the applications we tested. A basic problem with all these proposed functionals is their unreasonably large ultra-nonlocality. We discuss this issue in section V of this paper.

For actual implementation of the resulting non-local space-time theory we developed the following scheme. We rely on a standard spatial representation using a plane waves basis and pseudopotentials[35]. For the time propagation, we developed the Memory Replica technique, which allows us to use any ordinary differential equation propagator. In the present case, we employed the 5$^{th}$ order adaptive step-size Runge Kutta method[36].

The resulting method is applied to several setups involving laser – metal-cluster interaction. We first study line broadening of a surface plasmon in gold clusters, modeled by spherical Jellium sphere. We find the plasmon absorption line is considerably broadened and slightly shifted to the blue. The broadening is due to decay of the collective excitation into many low energy electron-hole pairs. In linear response regime, such a decay process cannot be accounted for by adiabatic functionals, which are able to describe only decay by single electron-hole pair excitation (or Landau damping, as it is sometimes called[19]). Next, we study second order phenomena associated with a two pulse experiment on such gold clusters. The pump is tuned to excite a surface plasmon of the cluster and the probe, given after delay $\tau$ is a short pulse which checks a wide range of frequencies. We focus on 2$^{nd}$ order dynamics which amount to absorption of a photon by the pump and then either absorption of a second photon by the probe or emission of a photon induced by the probe. The first pulse stes the system in coherent motion and the second pulse can check how this coherence is maintained as a function of time. We find that memory effects cause a significant damping of the coherences set up by the first pulse. The third example includes very strong fields where high harmonic generation and ionization is important. In contrast to the absorption spectrum, the HHG lines are weaker and considerably sharpened due to memory effects, when compared to ALDA.

We present the functional and potentials in section II. Then the time propagation method is explained in section III. The applications and results are described in section IV. A summary and discussion follows in section V.

## II.   M-MOMENT METHOD

We discuss in this section a relatively simple method for obtaining TC memory potentials in TDDFT. The method generalizes the center of mass method proposed in ref.[37]. It is based on a moment functional of the density which is a reference point in the electron density to which any observer can relate his memory terms. When this idea is implemented via an appropriate action functional which is translationally invariant. Thus, the resulting potentials are automatically TC and obey the zero force condition (i.e. obey Eq (1.3) and (1.5)).

## A. Translational invariance: notions and definitions

One technique for producing TC XC potentials that have zero net force is to set up a TI "action functional" of the density from which the XC potentials are derived. To explain what is a TI functional, consider a system of electrons and its two description by two moving observers. One observer describes a given point in 3D space as $\mathbf{R}$ while the second observer denotes the same point as $\mathbf{R}'$. In general we have:

$$\mathbf{R} = \mathbf{R}' + \mathbf{x}(t), \qquad (2.1)$$

Here $\mathbf{x}(t)$ is independent of the point and describes the time-dependent displacement between the two observers. The two observers are studying the same physical system of electrons. The lab observer describes the electron number density by the function $n(\mathbf{R},t)$ and the second observer, describes the same density by the function $n'(\mathbf{R}',t)$. Obviously, the density at a given physical point and time must be the same for both observers, so the relation between the two functions is:

$$n'(\mathbf{R}',t) = n(\mathbf{R},t) = n(\mathbf{R}' + \mathbf{x}(t),t). \qquad (2.2)$$

(the same relation as Eq. (1.2)). Now consider a functional $S_{XC}[n]$ of the density. We say it is *translationally invariant* (TI), if both observers, when they use the functional obtain the same value for their respective density functions. So, $S_{XC}[n]$ is TI if

$$S_{XC}[n'] = S_{XC}[n]. \qquad (2.3)$$

Thus $S_{XC}$ is TI if it yields the same result for the same physical system, irrespective of the observer. Now, if the exchange-correlation potential is the functional derivative of a TI functional,

$$v_{XC}(\mathbf{R},t) = \frac{\delta S_{XC}}{\delta n(\mathbf{R},t)}, \qquad (2.4)$$

then it obeys Eqs. (1.3) and (1.5) i.e. it is TC and it complies with the zero force condition[27]. Note that convincing arguments have been raised[38] that a relation encapsulated in Eq. (2.4) can not hold in general, because unless $S_{XC}$ is local in time it "violates" causality, i.e. leads to a dependence of $v_{XC}$ at time $t$ on the density at a later time. It was however shown that a TI action can still be obtained if one considers a mathematical device called the Keldysh contour[34].

The ALDA XC potential is derivable from a TI action:

$$S_{ALDA}[n] = \int_0^{t_f} dt \, E_{LDA}[n(t)] \qquad (2.5)$$

Where $E_{LDA}[n(t)]$ is the LDA energy associated with the density at a time $t$. Knowing that ALDA gives reasonable results, we are motivated to write the ALDA+M action functional as:

$$S_{ALDA+M}[n] = S_{ALDA}[n] + S_{mem}[n] \qquad (2.6)$$

In subsection B we build a simple TI $S_{mem}[n]$.

## B. Density M-moment Method

Now, assume that some functional $s_{mem}[n]$ from which the potential is to be derived is given. Most likely, this functional will not be TI to start with, so we enforce TI upon it by using the M-moment method described now. Consider the density moment frame, denoted:

$$\mathbf{r} = \mathbf{R} - \mathbf{D}(t). \qquad (2.7)$$

Where the $\mathbf{D}$ is the M-moment, a functional of $n$ is defined as:

$$\mathbf{D}[n](t) = \frac{\int \mathbf{R} M[n(\mathbf{R},t)] d^3 R}{\int M[n(\mathbf{R},t)] d^3 R} \equiv \frac{\int \mathbf{R} M[n(\mathbf{R},t)] d^3 R}{Q[n](t)}. \qquad (2.8)$$

$M(n)$ is the M-function: arbitrary except that it is positive and obeys $\lim_{n \to 0} M(n) = 0$. The simplest M-function is $M_{CM}(n) = n$, where $\mathbf{D}$ is simply center of mass $\mathbf{R}_{CM}$ of the electron distribution. However, we are not limited to this. We can take other M-functions, for example $M(n) = ne^{-\gamma n}$, $M(n) = n^2$ or $M(n) = 1 - e^{-\gamma n}$. No matter what choice of $M(n)$ we take, the M-function has the following transformation rule under changing to an accelerated frame:

$$\mathbf{D}[n'](t) = \mathbf{D}[n](t) - \mathbf{x}(t) \qquad (2.9)$$

Furthermore, the functional derivative with respect to the density yields:

$$\frac{\delta \mathbf{D}[n](\bar{t})}{\delta n(\mathbf{R},t)} = \frac{M'(\mathbf{R},t)}{Q[n](t)}[\mathbf{R} - \mathbf{D}[n](t)]\delta(t - \bar{t}). \qquad (2.10)$$

An observer in the M-moment frame will use a function $N(\mathbf{r},t)$ to describe the density, where:

$$N[n](\mathbf{r},t) = n(\mathbf{R},t) = n(\mathbf{r} + \mathbf{D}[n](t),t). \qquad (2.11)$$

It is straightforward, using (2.9), to verify that this density is actually TI:

$$\begin{aligned} N[n'](\mathbf{r},t) &= n'(\mathbf{r} + \mathbf{D}[n'](t),t) \\ &= n(\mathbf{r} + \mathbf{D}[n'](t) + \mathbf{x}(t),t) \\ &= n(\mathbf{r} + \mathbf{D}[n](t),t) \\ &= N[n](\mathbf{r},t) \end{aligned} \qquad (2.12)$$

From Eqs. (2.10) and (2.11), the functional derivative:

$$\frac{\delta N(\mathbf{r},\bar{t})}{\delta n(\mathbf{R},t)} = \delta(t - \bar{t})\Big[\delta(\mathbf{r} + \mathbf{D}(t) - \mathbf{R}) \\ + \frac{M'(n(\mathbf{R},t))}{Q(t)} \nabla n(\mathbf{r} + \mathbf{D}(t),t) \cdot (\mathbf{R} - \mathbf{D}(t))\Big]. \qquad (2.13)$$

We now relate the action $s_{mem}$ only to the M-moment system,

by defining:
$$S_{mem}[n] = s_{mem}[N[n]] \quad (2.14)$$

Since $N[n]$ is TI, we immediately see that $S_{XC}[n]$ is itself TI. Using (2.10), (2.13) and a bit of chain rule differentiation, we can derive the general form of the XC potential:

$$v_{mem}[n](\mathbf{R},t) = V_{mem}[N](\mathbf{R}-\mathbf{D}(t)) \\ + \mathbf{E}_{mem}(t) \cdot (\mathbf{R}-\mathbf{D}(t))M'[n(\mathbf{R},t)] \quad (2.15)$$

Where:
$$V_{mem}(\mathbf{r},\bar{t}) = \frac{\delta s_{mem}[N]}{\delta N(\mathbf{r},\bar{t})} \quad (2.16)$$

and:
$$\mathbf{E}_{mem}(t) = \frac{1}{Q(t)} \int V_{mem}(\mathbf{r},t) \nabla n(\mathbf{r}+\mathbf{D}(t),t) d^3 r \quad (2.17)$$

Since $S_{mem}$ is TI, we are assured that $v_{mem}$ is TC and leads to zero XC-force. This can also be checked directly on the final result. Note that in Eq. (2.15) $v_{mem}$ is a sum of two terms and that each of them is TC. Thus the second term is not needed for TC but its presence is required to ensure that the total force is zero as well.

In the rest of the paper, we specialize to the center of mass method, where $M(n) = n$. This choice of the M-function leads to a similar theory proposed in ref [27], however, as noted above our potential has the added benefit of having zero net force.

### C. Building the TC potential

We now choose a specific form for $s_{XC}[n]$, based on a given parameterization of the density response properties of the HEG, are encapsulated in a causal kernel $F(n,t)$, with:
$$F(n,\tau < 0) = 0 \quad (2.18)$$

There are several parameterizations of the kernel in the literature[21, 22, 30]. We concentrate on the general approach here. We define the Keldysh pseudo time variable $\tau$, where $\tau \in [0,\tau_f]$, and a mapping $t(\tau)$ of this variable onto physical time. The mapping is constrained by: $t(0) = t(\tau_f)$. More details can be found in ref. [34]. The XC action is thus:

$$s_{mem}[N] = \int d\mathbf{r}' \int_0^{\tau_f} \dot{t}(\tau') d\tau' \times \\ \int_0^{\tau'} d\tau'' F(N(\mathbf{r}',\tau'),\tau'-\tau'') \dot{N}(\mathbf{r}',\tau'') \\ (2.19)$$

From which, using the special properties of the Keldysh contour:

$$V_{mem}[N](\mathbf{r},\bar{t}) = \int_0^{\bar{\tau}} F'_{mem}(N(\mathbf{r},\bar{t}),\bar{t}-t'') \dot{N}(\mathbf{r},t'') dt'', (2.20)$$

where:
$$F'_{mem}(N(\mathbf{r}',t),t-t') = \left.\frac{\partial F_{mem}(n,t-t')}{\partial n}\right|_{n=N(\mathbf{r}',t)} \quad (2.21)$$

The XC potential of Eq. (2.15) becomes:
$$v_{mem}[n](\mathbf{R},t) = \int_0^t dt' F'_{mem}(n(\mathbf{R},t),t-t') \\ \times \partial_{t'} n(\mathbf{R}-[\mathbf{D}(t)-\mathbf{D}(t')],t') \quad (2.22) \\ + \mathbf{E}_{mem}(t) \cdot (\mathbf{R}-\mathbf{D}(t))$$

With a homogeneous XC "electric" field given by:
$$\mathbf{E}_{mem}[N](t) = \frac{1}{Q(t)} \int d^3 r \int_0^t dt' F'_{mem}(N(\mathbf{r},t),t-t') \times \\ \dot{N}(\mathbf{r},t')[\nabla N](\mathbf{r},t) \quad (2.23)$$

The form of the kernel function is obtained from the HEG dynamical linear response properties[26]. We can connect our result to this limit by developing our functional to first order around a homogeneous (space independent) gas density $n_0$. Note that in this limit the electric field is a second-order quantity. After some manipulations, we can show:

$$\tilde{F}'_{mem}(n,\omega) = \frac{f^h_{mem,L}(n,\omega)}{-i\omega}, \quad (2.24)$$

Where $f^h_{mem,L}(n,\omega)$ is the longitudinal linear response kernel of the HEG[21, 22].

Summarizing, the total ALDA+M XC potential we use is:
$$v_{ALDA+M}(\mathbf{R},t) = v_{ALDA}(\mathbf{R},t) + \\ \int_0^t dt' F'_{mem}(n(\mathbf{R},t),t-t') \partial_{t'} N(\mathbf{R}-\mathbf{D}(t),t') \quad (2.25) \\ + \mathbf{E}_{mem}(t) \cdot (\mathbf{R}-\mathbf{D}(t)),$$

## III. PROPAGATION: THE REPLICA METHOD

Our numerical representation is based on a standard plane-waves basis set method[35] with image screening[39]. The method allows accurate spatial derivative calculations and yields accurate interpolations for calculating $N(\mathbf{r},t) = n(\mathbf{r}+\mathbf{D}(t),t)$.

In ALDA applications, the plane waves basis combined with the 5th order adaptive time step Runge-Kutta method[40] for propagating the TDKS equation is an efficient and accurate scheme. However, when using memory functionals the propagation method cannot be used. To understand why, we note that the RK method propagates ordinary differential equations of the type:

$$\dot{y}(t) = f(y(t),t) \quad (3.1)$$

This form is compatible with the TDKS equations of ALDA,

where $y$ stands in for the set $\{\psi_n\}_{n=1}^{N_e}$ KS orbitals and $f$ for $-iH_{KS}\psi_n$. However, with a memory term present, the TDKS equations are of the form:

$$\dot{y}(t) = f\left(y(t), y(t-\Delta t), y(t-2\Delta t),...,t\right) \quad (3.2)$$

Where $\Delta t$ is some (approximate) coarsening. This is not of the type to which efficient RK methods are applicable.

In order to convert Eq. (3.2) to the type (3.1), we define *chronological* replicas:

$$y_n(t) \equiv y(t-n\Delta t), \qquad n = 0,1,... \quad (3.3)$$

Clearly:

$$\begin{aligned}\dot{y}_n(t) &= \dot{y}(t-n\Delta t)\\ &= f\left(y(t-n\Delta t), y(t-(n+1)\Delta t),...,t-n\Delta t\right) \quad (3.4)\\ &= f\left(y_n(t), y_{n+1}(t),...,t-n\Delta t\right)\end{aligned}$$

Defining

$$f_n(y_0, y_1,...,t) \equiv f(y_n, y_{n+1},...,t-n\Delta t) \quad (3.5)$$

It is possible to write, in vector notation:

$$\dot{\mathbf{y}}(t) = \mathbf{f}(\mathbf{y}(t), t) \quad (3.6)$$

We make an additional approximation, by keeping only a finite number $N_r$ of replicas, so that the relevant history extends backward to time $t-T_m$ (with $T_m = N_r \Delta t$). Eq. (3.6) is now of the type (3.1), i.e. amenable to RK propagation. The price to be paid here is that of propagating and keeping $N_r$ replicas of the KS orbitals. Thus, the method introduces two additional parameters: the number of history steps $N_r$ and the coarsening time step $\Delta t$. Convergence tests of the final results with respect to the limits $T_m \to \infty$ and $\Delta t \to 0$ should be made. In the calculations we present here we used $T_m = 6\,au$ and $\Delta t = 0.75\,au$. We have checked that these values give reasonably converged results.

## IV. APPLICATIONS

Our 3D simulations focus on small spherical metal gold clusters. We model such systems using a spherical Jellium model, where the ionic charge of gold is smeared to its average value ($n = \left[\frac{4\pi}{3}r_s^3\right]^{-1}$ with $r_s = 3a_0$) of an appropriate radius. Please note our nomenclature: we use the acronym "$Au_8$ spherical Jellium cluster" to denote a Jellium sphere which contains 8 s electrons of gold (neglecting the d-electrons) and the total positive charge is 8 as well.

The system is placed in a box of dimension $L_x \times L_y \times L_z$ the grid spacing is uniform $\Delta x$ and the number of gridpoints or plane-waves is $N_{x,y,z}\Delta x = L_{x,y,z}$.

For each application we present results of ALDA and the memory functional. The memory parameterization we used in this work is due to and Iwamoto and Gross i.e.

$$\text{Im}\, f_{XC} = \frac{a(n)\,\omega}{\left(1 + b(n)\,\omega^2\right)^{5/4}}, \quad (4.1)$$

Where $a(n)$ and $b(n)$ are known functions of the density[21, 22], see Appendix A for details. For numerical stability we smoothly truncate the kernel so it is zero when the density is very small ($r_s > 6$).

### A. Long wavelength linear response of HEG

Our first example is made mainly for demonstration, since the functional is designed to closely imitate what we know about the response of the HEG to long wavelength sinusoidal perturbations. We consider a homogeneous electron gas of density parameter $r_s = 3$ (corresponding to the density of gold). A box of volume $V = L^3$ is setup and a plane-waves basis is used with periodicity in three dimensions. In the present calculations we choose $L = 16.3a_0$ and the number of electrons is $N_e = 38$. This electron number gives a closed shell system. The Kohn-Sham orbitals are the plane waves $\psi_\mathbf{k}(\mathbf{r}) = e^{i\mathbf{k}\cdot\mathbf{r}}$ where $k_x$ $k_y$ and $k_z$ are integer multiples of $\frac{2\pi}{L}$ and the $N_e/2$ states with the lowest $|\mathbf{k}|$ are occupied. At $t > 0$ the system is perturbed by a very short Gaussian pulse coupled by a long-wavelength field:

$$v_{ext}(\mathbf{r},t) = E_0 Z_k(\mathbf{r}) f(t) \quad (4.2)$$

Here $E_0$ is a weak enough field so that linear response is dominant (we choose $E_0 = 10^{-3} E_h \left[ea_0\right]^{-1}$) the spatial and temporal forms of the perturbation are:

$$Z_k(\mathbf{r}) = \frac{\sin kz}{k}; \quad f(t) = e^{-(t-t_0)^2/2\sigma^2} \quad (4.3)$$

$k = \frac{2\pi}{L}$. The pulse shape is a very short Gaussian:

$$f(t) = e^{-(t-t_0)^2/2\sigma^2} \quad (4.4)$$

In Eq. (4.2), $t_0 = 8\,\hbar E_h^{-1}$ $\sigma_0 = 2\,\hbar E_h^{-1}$. The reason we use $Z_k$ and not a dipole field is that for the HEG and a dipole field there are no memory effects because of the Harmonic Potential Theorem[23].

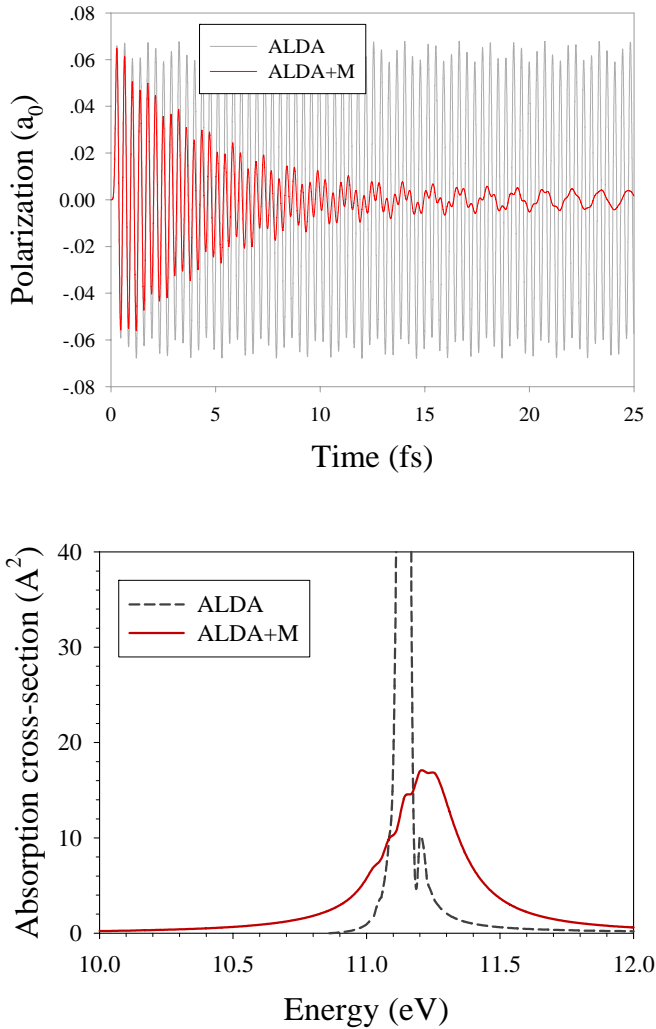

Figure 1: The Z-Z correlation-function vs time for the finite HEG slab (panel (a)). The corresponding plasmon absorption cross section lineshape (panel (b)). This figure compares ALDA with ALDA+M.

The simulation is done in real time, thus we start from the ground-state Kohn-Sham orbitals (which are in this case the lowest $N_e/2$ plane-waves) with density $n(\mathbf{r},0) = n_0$, constant throughout space. We apply the time-dependent external potential (4.2) for $t > 0$ and using the Kohn-Sham equations, we evolve the electronic density $n(\mathbf{r},t)$ in time. The time-dependent expectation value of $Z_k$,

$$\left\langle \hat{Z}_k \right\rangle_t = \int \left[ n(\mathbf{r},t) - n_0(\mathbf{r}) \right] Z_k(\mathbf{r}) d^3r \qquad (4.5)$$

Is recorded at equal intervals. It can be shown[41], (see also discussion in ref. [42]) that:

$$\left\langle \hat{Z}_k \right\rangle_t = \frac{E_0}{i\hbar} \int_0^\infty f(t-\tau) \left\langle \psi_{gs} \left| \left[ \hat{Z}_k(\tau), \hat{Z}_k(0) \right] \right| \psi_{gs} \right\rangle d\tau, \quad (4.6)$$

where $\psi_{gs}$ is the many-body ground-state. Thus, $\left\langle \hat{Z}_k(t) \right\rangle$ is proportional to the imaginary part of the correlation function $\left\langle \hat{Z}_k(t) \hat{Z}_k(0) \right\rangle$ (we refer to this below as the "Z-Z correlation function") which is related to the energy absorption, just like in dipole excitation (see for example ref. [43]).

The Z-Z correlation-function in real time is plotted in Figure 1. We show two correlation-functions computed using ALDA and the ALDA + memory functional in real-time and frequency. The difference is striking. While the correlation function computed via ALDA shows no observable damping on the timescale of tens of fsec, the high frequency part of the ALDA+M transient is strongly damped within about 5-10 fsec. Only a low-frequency mode survives this strong decay. The high frequency line is the bulk plasmon and is shown in panel (b) of Figure 1. Two features are noticeable: the memory functional shifts the maximal absorption to the blue by about 0.1 eV and the line-width is about 0.5 eV. For ALDA the plasmon line-width is artificial, i.e. just reflects the finite propagation time. The width of the plasmon line we find is in good agreement with theoretical estimates[24].

### B. Absorption spectrum of Au$_{18}$ spherical Jellium cluster

In this section, we study the absorption spectrum of a finite size spherical Au cluster, focusing on the surface plasmon excitation. This observable is the subject of many recent studies [44-47]. Experiments have shown[44, 48, 49] that the overall line width of the surface plasmon in clusters and nanospheres is large, on the order of 0.5-1eV.

We use the popular spherical jellium model, by which only the s-electrons (one per Au ion) is considered and the ionic charge is smeared (almost) uniformly on a sphere, namely the spatial charge density at point $\mathbf{r}$ is:

$$n_+(\mathbf{r}) = \frac{n_0}{1 + e^{[2r - D_0]/\lambda}}. \qquad (4.7)$$

The diameter of our spherical cluster is $D_0 = \sim 0.83$nm and it contains a total positive charge of $18e$. Within the sphere, the charge density is almost uniform, at the average ionic density of gold. The smeared charge near the surface is smoothly cut off ($\lambda$ is a smoothing parameter), so the density is zero outside of the sphere. The sphere is neutral so there are 18 electrons in the system, namely the Au valence s-shell electrons. Even with this small number of electrons one can usefully discuss excitation modes such as collective (plasmon) oscilations[50]. We designate this and similar model metallic clusters as "spherical jellium cluster Au$_n$" where $n$ is the positive charge of the sphere (in units of $e$).

The photoabsorption cross-section of Au$_{18}$ spherical Jellium cluster is computed using the dipole-dipole correlation function, as described in[51]. Basically we use the method of the previous subsection (Eqs. (4.3)- (4.6)) but with $\hat{Z}_k$ replaced by the dipole operator $\hat{z}$. The diameter of the Au$_{18}$ cluster is 0.83nm. The dipole correlation function and spectrum is presented in Figure 2. The two calculated signals, one based on ALDA and the other on memory functional, are very different.

The most notable difference in the time domains is the decay of the dipole-dipole correlation function on a time scale of 30 fs when memory functionals are used. Examination of the surface plasmon line at ~3.6 eV shows a small blue shift and a large 0.1 eV line width. It should be noted that the ALDA line-width is artificial resulting solely from the finite propagation time.

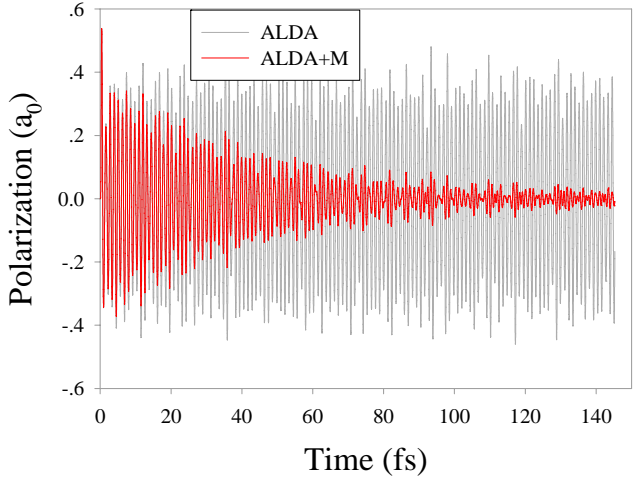

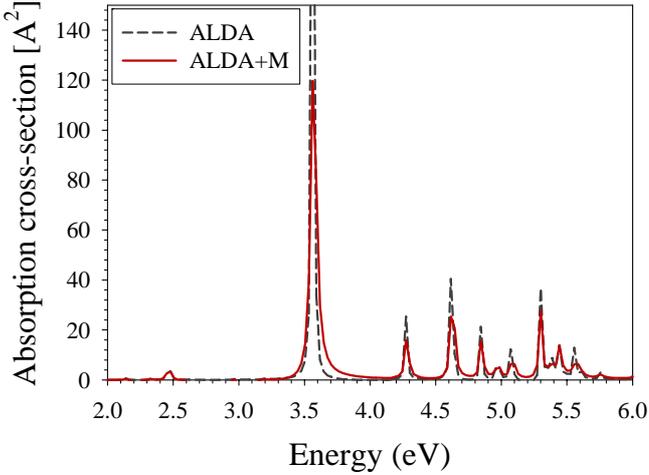

Figure 2: The polarization vs. time for the Au$_{18}$ spherical jellium cluster (top panel) after the short pulse excitation. The corresponding absorption cross sections are shown in the bottom panel.

**C. Plasmon decay: a pump-probe 2D spectroscopy study**

Pump-probe experiments on metals form a basic suit of methods with which the lifetime of electronic excitations can be studied[52-59]. In this section, we use a 2-pulse setup to pump (excite) and probe the plasmon decay in a small Au$_8$ spherical jellium cluster. Our setup is as follows: the pump has a reasonably well-defined frequency $\omega_p$ peaking at time $t_0$; and the probe is of extremely short duration (tens of atto-second), and contains a wide range of frequencies. The probe is given within a time delay $\tau$ relative to the pump. To be specific, we consider the following coherent pulse shape, which contains the pump, of strength $E_0$ and probe, of strength $E_1$:

$$p(t;\tau,E_0,E_1) = E_0 e^{-\frac{[t-t_0]^2}{2\Sigma^2}} \sin\omega_p t + E_1 e^{-\frac{[t-(t_0+\tau)]^2}{2\sigma^2}} \quad (4.8)$$

The parameters we use are collected in Table 1.

Table 1: Pump-probe parameter for Eq. (4.8). All quantities in atomic units.

| | | |
|---|---|---|
| $t_0$ | Time of pump | 200 |
| $\Sigma$ | Duration of pump | 80 |
| $\omega_p$ | Frequency of pump | 0.13 |
| $E_0$ | Field of pump | 0.001 |
| $\tau$ | Pump-probe delay | Variable: 200, 600, 1000, 1400, 1800 |
| $\sigma$ | Duration of probe | 2 |
| $E_1$ | Field of probe | 0.001 |

Note that the first pulse is of relatively well-defined frequency $\omega_p$, exciting the surface plasmon (at energy ~3.5 eV as seen in Figure 2 for a different cluster, but the plasmon frequency is similar) while the second pulse is of extremely short duration, containing a wide spectrum of frequencies. The electric field couples to the electronic dipole operator. A TDDFT run is then performed, starting from the ground-state and a dipole signal $S(t;\mathbf{E})$ is recorded. When the electric fields of both pulses are small we can write (all other parameters are held constant):

$$S(t;\tau,\mathbf{E}) = S_0(t;\tau) + \mathbf{b}(t;\tau)^T \mathbf{E} + \frac{1}{2}\mathbf{E}^T \overleftrightarrow{\mathbf{L}}(t;\tau) \mathbf{E} + ... (4.9)$$

$S_0$ is in principle zero, but for numerical calculations attains nonzero values due to numerical noise. The coefficients $\mathbf{b}(t;\tau) \equiv (b_0, b_1)$ are linear response spectra while the 3 elements of the 2×2 symmetric matrix $\overleftrightarrow{\mathbf{L}}(t;\tau)$ are spectra for 2$^{nd}$ order processes. In particular, the element $L_{01}(t;\tau)$ corresponds to second order processes which are linear in $E_0$ and $E_1$. Thus $L_{01}$ describes absorption of a photon $\sim \omega_p$ by the pump and a subsequent absorption or induced emission of a second photon at frequency $\omega_2$. The final state of the cluster after this process is at energy $\hbar(\omega_p + \omega_2)$.

In order to obtain the spectrum for this type of process we need to take the second derivative $L_{01}(t;\tau) = \left.\frac{\partial^2 S(t;\tau,\mathbf{E})}{\partial E_0 \partial E_1}\right|_{\mathbf{E}=0}$, which we approximate as:

$$L_{01}(t;\tau) \approx \frac{1}{4E_0 E_1}\big[S(E_0,E_1) - S(-E_0,E_1) \\ - S(E_0,-E_1) + S(-E_0,-E_1)\big] \quad (4.10)$$

Thus, we need 4 separate TDDFT dipole signals runs from which the signal $L_{01}(t;\tau)$ can be obtained from each time delay $\tau$. This is in effect a 2-dimensional spectrum of the system.

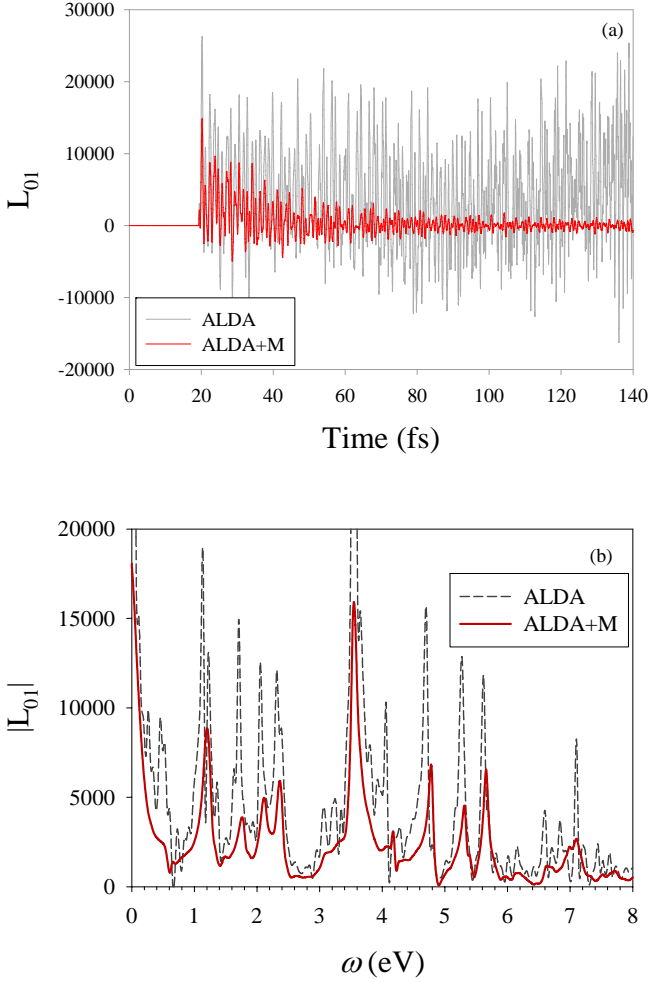

Figure 3: Memory effects on the real-time (a) and frequency (b) 2-photon response $L_{01}$ for $Au_8$ spherical jellium cluster at $\tau = 25\,fs$.

One such a signal, for $\tau = 1000\,atu$ (about 25 fs) is shown for the $Au_8$ cluster in Figure 3. The $L_{01}$ signal starts after 25fs. In Figure 3a we see that the $L_{01}$ ALDA and ALDA+M signals are very different, as the latter is quickly damped, while the first seems to oscillate indefinitely (on the time scale of the calculation). In the frequency domain, Figure 3b, we observe that the sharp ALDA spectral features are either absent or much reduced in the ALDA+M spectra. Furthermore, some of the peaks are slightly blue-shifted

Since the first pulse is also of short duration, it excites many modes in the electron gas. This creates a linear combination of vibrating modes in the non-homogeneous electron gas of the metal cluster. The measured effect of the second pulse will thus depend on the many frequency *differences* that exist between these modes. The way this "vibrational coherence" is preserved in time is an important probe of the dephasing and relaxation processes in the metal cluster.

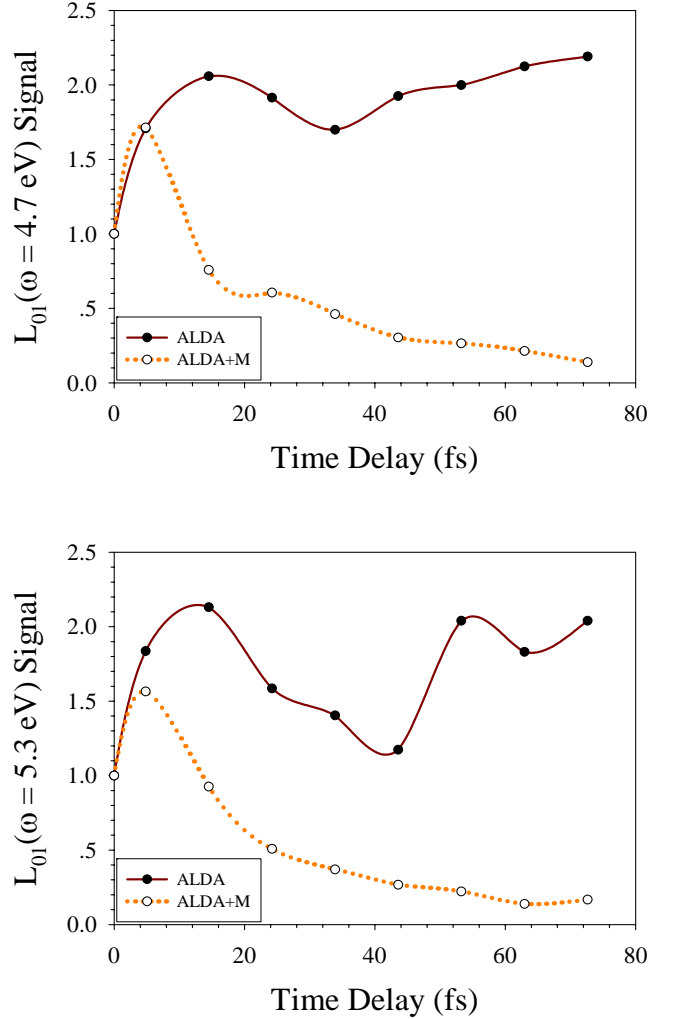

Figure 4: The ALDA and ALDA+M pump-probe $L_{01}$ 2-photon absorption signals in $Au_8$. The first photon is a 3.6 (plasmon) frequency and absorbed from the pump while the second is photon is obtained from the probe, 0.9 eV (top) and 1.7 eV (bottom).

In Figure 4, we examine the strength of the process where the absorption of a photon from the pump excites the plasmon, followed by absorption of an additional photon, from the probe. Two lines are selected as an example, the $\omega = 4.7\,eV$ and $\omega = 5.3\,eV$. In the ALDA calculation the yield is initially grows with time delay but then becomes more or less constant at around 2.0 (the units here are arbitrary). In the ALDA+M case however, the yield initially goes up (as in the ALDA case) but then almost monotonically goes down. This shows that the population of the plasmon decays as a function of time. The decay is very similar in both cases. We may say that the decay shows two types of behavior. There is a fast decay of about 10 fsec, followed by a slow decay of about 30-40 fs.

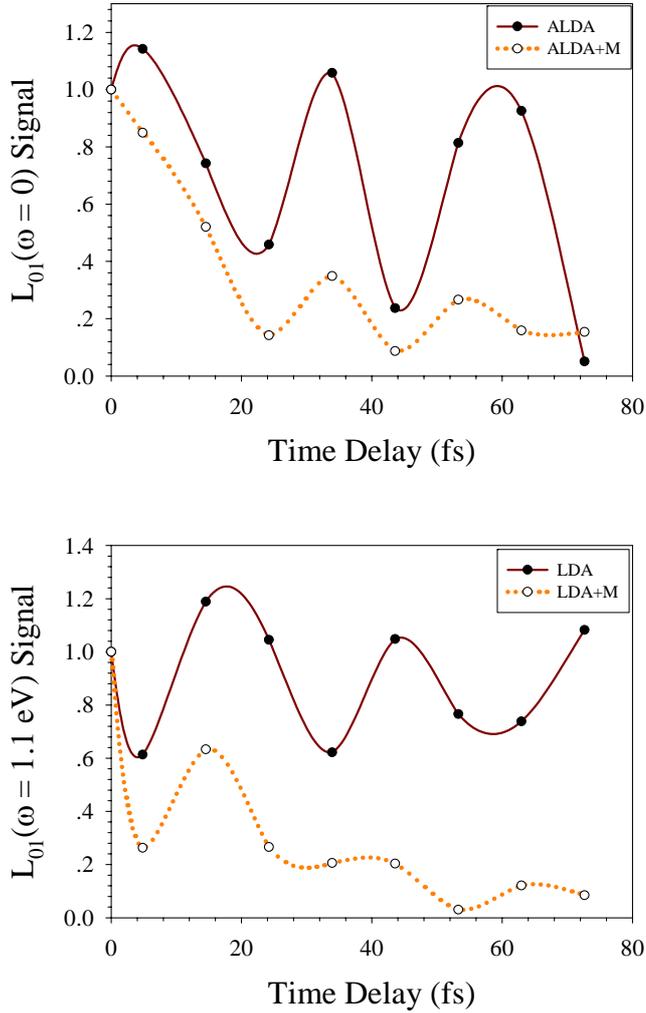

Figure 5: The ALDA and ALDA+M pump-probe $L_{01}$ 2-photon absorption-emission signals in Au$_8$. Absorption of the pump photon excites the plasmon at 3.5eV. The probe stimulates emission of a second photon at 3.6 eV (top) and 2.5 eV (bottom).

It is now interesting to study the 2-photon signal in the case where the first photon is absorbed but the second photon is *emitted*. Thus we look at two transition $\omega = 0$ and $\omega = 1.1 eV$. The absorption-emission spectrum is more oscillatory than the absorption-absorption spectrum. The oscillations are manifestations of coherences which exist between different modes with nearly the same frequency. In Figure 5, we see that the ALDA+M transients in exhibits, once again, damped coherence oscillations when compared to the ALDA transients. As in the absorption-absorption case, it seems that there are two time scales for damping: a fats and a slower one.

### D. High Harmonic generation in Au$_{18}$

In recent years, TDDFT within the ALDA were used to study high harmonic generation (HHG) in molecules and clusters[16, 60-63]. Since these TDDFT applications have not taken onto account memory effects, we investigate this issue here, in the context of metal clusters. We expose a spherical-Jellium Au$_{18}$ cluster to a short (50 fs) pulse of intense laser radiation at a frequency $\omega_0$. The emission spectrum is associated with the dipole acceleration, given by:

$$P \propto \left| \int_0^\infty \ddot{d}(t) e^{-i\omega t} dt \right|^2 \quad (4.11)$$

We have excited the system using a laser pulse electric field profile given by:

$$E_z(t) = E_0 \sin^2\left[\frac{\pi t}{T}\right] \cos \omega_0 t \quad (4.12)$$

In the example we study here, we took pulse duration is $T = 2000\hbar E_h^{-1}$ (about 50 fs) and frequency $\omega_0 = 0.1 E_h \hbar^{-1}$ and electric-field $E_0 = 0.01 E_h (ea_0)^{-1}$. At these intensities there is appreciable ionization. In order to account for electron flux moving away from the cluster, never to return, we impose absorbing boundary conditions using a negative imaginary potential[64] (NIP) placed asymptotically in the direction of the electric field polarization[60]. This potential removes the electron density that gets pushed out and away from the cluster. The functional form of the potential is:

$$W(\mathbf{r}) = -iA \times \theta(|z|-a) \times (|z|-a)^3 \quad (4.13)$$

With:

$$A = 0.00064 E_h a_0^{-3} \qquad a = 15 a_0 \quad (4.14)$$

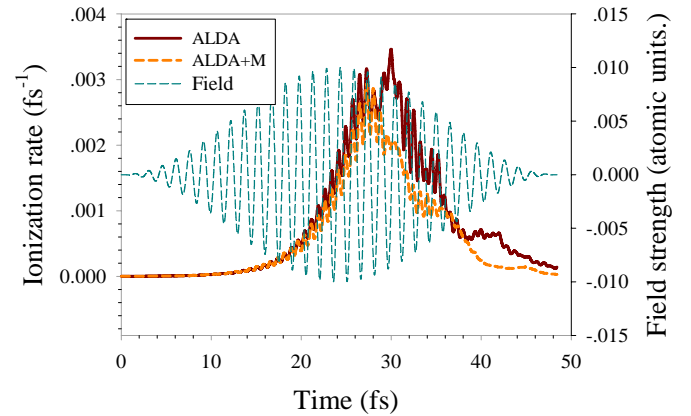

Figure 6: Memory effects in the transient ionization rate for the Au$_{18}$ spherical jellium cluster during a short and intense laser pulse.

Let us first discuss ionization. The effective rate of ionization is defined as the rate of loss of electron density within the simulation box. In our calculation, charge that is very far from the cluster is absorbed by the negative imaginary potential. These calculated rates are compared in Figure 6. One sees from the figure that the ALDA rate has much more structure than the corresponding transient of ALDA+M. In particular, the ALDA transient has three peaks main peaks, at 27, 32 and 35 fs while the ALDA+M peaks only once, at 27 fs (immediately after the laser electric field passes its maximum). This is in compliance with the previously seen tendency of the memory effects to damp oscillations[19].

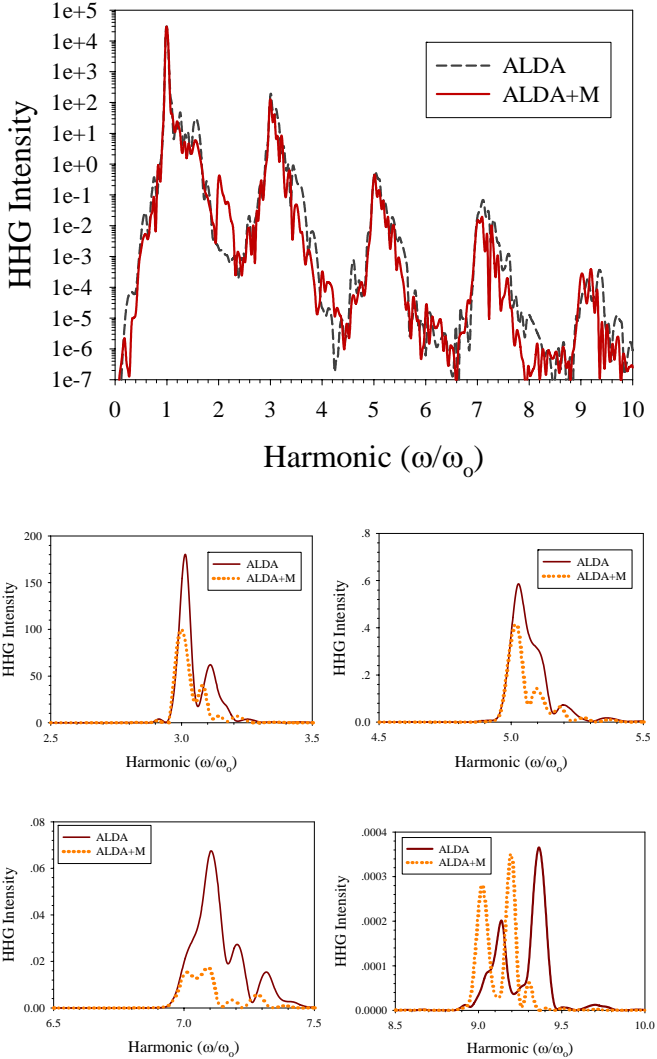

Figure 7: Memory effects on the HHG spectra for the $Au_{18}$ spherical jellium cluster. The top panel shows the spectrum on a log scale. Close-ups on the line shapes in linear scale of harmonics 3,5,7,9 are shown in the bottom panels.

Next we study the HHG spectra. Due to the spherical symmetry of our system, only odd harmonics can be formed for a continuous wave (CW) field[65]. However, since we used a short pulse, the pulses emission lines have a complicated shape although the peaks are centered near (but not exactly at) the odd integer harmonics. This is shown in Figure 7. The memory effects at the odd harmonics serve to reduce the HHG intensity by a factor of 1.5-3. The memory functional spectra is weaker but more concentrated on the odd-integer harmonics. Features which are a result of the finite pulse are washed away so the harmonic peaks are relatively enhanced. An interesting phenomenon, somewhat in contradiction to the behavior at odd harmonics, is that the 2$^{nd}$ harmonic, although very weak, is considerably stronger in the ALDA+M calculation that the ALDA one.

## V.  SUMMARY AND CONCLUSIONS

In this paper we developed a memory action functional which is translationally invariant. The potential we derive from this action is translationally covariant (similar in spirit to that of Vignale[27]) and has the additional merit of obeying the zero net force condition. Both conditions are automatically obtained when the potential is derived from a TI action. The memory functional is added to the ALDA functional, resulting in a new functional we dub ALDA+M. The M part of the functional depends on a parameterization of the HEG kernel, which we took to be the Iwamoto-Gross-Kohn[21, 22] functional.

We then described how memory potentials are handled numerically within Runge-Kutta propagation method. The resulting theory and numerical method is then applied to a small set of examples involving laser-metal cluster interaction. By examining the difference between various observables in the ALDA and ALDA+M calculations we learn what kinds of effects memory terms bring into the calculation. It has already been established[24] that memory effects in the linear response regime (or in the nearly homogeneous case) are largely viscous effects that damp the absorption lines, e.g. that of bulk plasmon. Beyond linear response, we find in this work that memory effects stay mostly viscous in nature.

We have studied the memory effects on absorption lines in small metal clusters. We found some absorption lines gain considerable width due to damping effects, associated with energy redistribution between the various modes of the metallic electron gas. However, the broadening we found was considerably less (by a factor 5-10) than experimental broadening for gold clusters[48] and nano crystals[44, 49]. We attribute this mainly to the inadequacy of the Jellium model. It is rigorously true that there are no memory broadening effects for electrons in a Harmonic trap excited by a homogeneous electric field[23]. Now, the external potential for an electron inside a Jellium sphere is exactly harmonic (as long as the electron is inside the sphere). Memory broadening effects are therefore due only to the anharmonicity on the surface. A more realistic model for metal clusters will have considerable anharmonicity inside the cluster as well and may lead to considerably larger broadening.

Ultra-fast pump-probe spectroscopy is often used to study decay processes in metals and we show that this is also useful as a computational tool the details of the plasmon decay are revealed by such settings. Once again, the ALDA calculation shows no damping while the ALDA+M transient exhibits double time-constant damping. The effect of memory on the high harmonic generation line shapes is interesting. Here it causes the line widths to *narrow*. This is probably a result of memory effectively damping all but the integer (odd) harmonic generation process.

The memory potential we derived here is a simple alternative to use of Lagrangian coordinates. However, such a potential can be severely criticized as being too ultra-nonlocal. Imagine a system composed of two molecules or clusters which are very far from each other. Now, excite, using a local time-dependent field just one of these systems. Because the other system is so far away, it should not be affected and the dynamics in the first system are too not affected by the presence of the second. However, since the M-moment is an intensive (as opposed to extensive) property of the entire system, it is obvious that the second system will be spuriously – through the memory. Conversely, the first system will behave differently in the presence of the second system, even when the two are far apart. Both features are non-physical and show that the M-moment approach is not a generally valid approximation, especially when large systems are involved. We say that the center of mass memory is not "size consistent". Future work is under development in order to solve this size consistency problem, without ruining the good features of the potential, i.e. its translational covariance and its compliance with the zero force condition.

**Acknowldegement**. We gratefully acknowledge support of the German Israel Foundation. RB thanks Sandy Ruhman, Ronnie Kosloff and Daniel Neuhauser for illuminating discussions on the subject of the paper.

## APPENDIX A  THE MEMORY KERNEL

The time domain kernel is related to the frequency-domain XC kernel of the HEG through (see Eq. (2.24)):

$$F'(n_0,t) = \frac{i}{2\pi}\int_{-\infty}^{\infty} d\omega \frac{f_{xcL}^h(n_0,\omega)}{\omega} e^{-i\omega t}, \quad (A.1)$$

We use the Iwamoto-Gross-Kohn functional given by[21, 22]

$$\mathrm{Im}\, f_{xcL}^h = \frac{a(n)\omega}{\left(1+b(n)\omega^2\right)^{5/4}} \quad (A.2)$$

(where $a(n)$, $b(n)$ are taken given in[21, 22]). The real part can be obtained from the Kramers-Kronig relation and after some manipulation[34]

$$F'(n_0,t) = -\frac{2}{\pi}\int_0^{\infty} \frac{\mathrm{Im}\, f_{xcL}^h(n_0,\omega)\cos\omega t}{\omega} d\omega \quad (A.3)$$

The integration can in (A.1) be done analytically yielding:

$$F'(n,t) = \frac{a(n)}{\sqrt{b(n)}}\phi_{3/4}\left(t/\sqrt{b(n)}\right) \quad (A.4)$$

Where the reduced kernel is:

$$\phi_u(x) = -\frac{2^{1/4}}{\sqrt{\pi}\,\Gamma(5/4)} x^u K_u(x) \quad (A.5)$$

$K_u(x)$ is the modified Bessel function of the second kind. We plot the reduced kernel $\phi_{3/4}(x)$ in Figure 8. One sees that the kernel is practically zero when $t \approx 5\sqrt{b(n)}$.

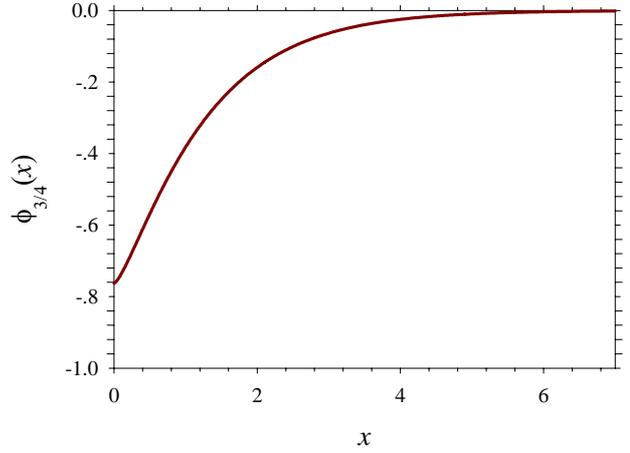

Figure 8: The reduced kernel $\phi_{3/4}(x)$.